\documentstyle[nato,numreferences,epsf,graphicx]{crckapb}

\begin{opening}
\title{Analytical study of low temperature phase 
of $3D$ LGT in the plaquette formulation}

\author{Oleg Borisenko, \ Sergei Voloshin }
\institute{Institute for Theoretical Physics, \  Ukrainian Academy of \\ 
Sciences, \ 03143 Kiev, Ukraine}
\author{Manfried Faber}
\institute{Atominstitut der \"osterreichischen Universit\"aten,\\
           Arbeitsgruppe Kernphysik, TU Wien, A--1040 Vienna, Austria}
\runningtitle{Analytical study of low-temperature phase of $3D$ LGT 
in the plaquette formulation}
\runningauthor{O.\ Borisenko, S.\ Voloshin, M.\ Faber }
\end{opening}
\begin{document}

\begin{abstract}
We develop an analytical approach for non-abelian gauge 
models within the plaquette representation where the plaquette
matrices play the role of the fundamental degrees of freedom.
We start from the original Batrouni formulation and show how it 
can be modified in such a way that each non-abelian Bianchi 
identity contains two connectors instead of four. Using this 
representation we construct the low-temperature expansion 
for $U(1)$ and $SU(N)$ models on a finite lattice and discuss 
its uniformity in the volume. Next, we derive a dual representation 
for the 't Hooft loop in the $SU(2)$ model and describe monopoles 
in the maximal axial gauge.  
\end{abstract}

\renewcommand{\thefootnote}{\fnsymbol{footnote}}
\footnotetext[0]{Presented by O.~Borisenko
at the NATO Advanced Research Workshop ``Confinement, Topology,
and other Non-Perturbative Aspects of QCD'',
January 21--27, 2002, Star\'a Lesn\'a, Slovakia.}
\renewcommand{\thefootnote}{\arabic{footnote}}

\section{Plaquette formulation}

Lattice gauge theory (LGT) can be formulated in many equivalent ways. 
The original Wilson formulation is given in terms 
of group valued matrices on links as fundamental degrees of 
freedom \cite{wilson}. The partition function reads
\begin{equation}
Z = \int DU \ \exp \{ -\beta S[U_{\mu}(x)] \} \ , 
\label{partfunc}
\end{equation}
\noindent
where $S$ is the standard Wilson action and the integral is calculated over 
the Haar measure on the group at every link of the lattice. 
Several years later the dual representation for abelian LGT's was 
constructed in \cite{dualu1}. Extensions to nonabelian groups have 
been proposed in ninetees in \cite{dualsun}-\cite{conf4}. 
A closely related approach is so-called
plaquette representation invented in the continuum theory by 
M.~Halpern \cite{halpern} and extended to the lattice models in \cite{plrepr}.
In this representation the plaquette matrices play the role of the dynamical 
degrees of freedom and satisfy certain constraints expressed through Bianchi 
identities.
Both dual and plaquette formulations are not so popular 
in the case of nonabelian models, probably due to their complexity.
In the nonabelian case above mentioned 
constraints appear to be highly-nonlocal 
and this fact makes an analytical study of the model rather difficult. 
In spite of this complexity the plaquette formulation has  
certain advantages over the standard one \cite{plrepr}:
duality transformations, Coulomb-gas representation, 
strong coupling expansion and mean-field approximation 
look more natural and simpler in the plaquette formulation. 
Nevertheless we believe the main advantage of this formulation
lies in its applications to the low-temperature region. 
Let $V_p$ be a plaquette matrix in $SU(N)$ LGT. 
The rigorous result of \cite{mackpetk} asserts that the
probability $p(\xi )$ that ${\mbox {Tr}}(I-V_p)\geq \xi$ is bounded by
\begin{equation}
p(\xi ) \leq {\cal O }(e^{-b\beta\xi}) \ , \ \beta\to\infty \ , \ 
b={\rm {const}} 
\label{chestplaq}
\end{equation}
\noindent
uniformly in the volume. Thus, all configurations with 
$\xi\geq O(\beta^{-1})$ are exponentially suppressed.
The Gibbs measure at large $\beta$ 
is strongly concentrated around configurations 
on which $V_p\approx I$. This property justifies expansion of plaquette 
matrices around unity when $\beta$ is sufficiently large.

This is our first motivation to construct a low-temperature
expansion of gauge models using the plaquette formulation.
The well-known problem of the standard perturbation theory (PT) 
can be shortly formulated as follows. When 
the volume is fixed, and in the complete axial gauge the link matrices perform 
small fluctuations around the unit matrix and the PT works very well producing the 
asymptotic expansion in inverse powers of $\beta$. However, 
in the thermodynamic limit (TL) system deviates arbitrarily far from 
the ordered state, no saddle point exists anymore and configurations of 
link matrices are distributed uniformly in the group space.
That there are problems with the conventional PT was proven in \cite{seiler}, 
where it was shown the PT results depend on the boundary conditions (BC) 
used to reach the TL. All this rises the question whether PT gives 
asymptotic expansion uniformly valid in the volume.  
Fortunately, even in the TL the plaquette matrices are close to unity,
the inequality (\ref{chestplaq}) holds and thus provides a basis for 
the construction of the low-temperature expansion.

Our next motivation is to derive the dual representation of nonabelian models 
which could be used for their analytical study. Here we formulate such 
representation for $SU(2)$ LGT and compute dual form of the 't Hooft loop. 
We also describe monopoles of the maximal axial gauge. 

\begin{figure}[ht]
\centerline{\epsfxsize=6cm \epsfbox{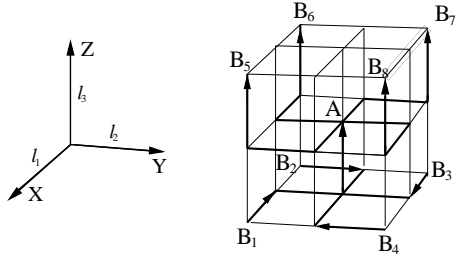}}
\caption{\label{connector} Structure of connectors on the lattice.}
\end{figure}

Now we outline the derivation of our plaquette formulation (all technical 
details will be given in a forthcoming publication \cite{plaq}).
Consider a $3D$ lattice with free BC on the link gauge matrices. 
We make a change of variables in the partition function (\ref{partfunc}) 
$V_p=U_{\mu} (x)U_{\nu} (x+\mu)U_{\mu}^{\dagger}
(x+\nu)U_{\nu}^{\dagger}(x)$. The partition function gets the form 
\begin{equation}
\label{func}
Z = \int \prod_{p} dV_{p} \exp \left[ \beta \sum_{p}
{\rm Re \ Tr} V_{p} \right] J(V_p) \ ,
\end{equation}
\noindent
\begin{equation}
J(V_p) = \int \prod_ldU_l \ \prod_p 
\delta \left ( V_p^{\dagger} \prod_{l\in p}U_l - I \right ) 
= \prod_cJ(V_c) \ .
\label{jacobdef}
\end{equation}
\noindent
The last equations are rather formal, for we have to specify 
the order of multiplication of nonabelian matrices. 
To do that we define in a cube a vertex $A$ and separated by the space 
diagonal a vertex $B$, see Fig.\ref{connector}. This assignment 
we extend for all neighbouring cubes and finally for the whole lattice. 
The plaquette matrices $V_p$ we insert at the vertices $A$ or $B$ of each 
plaquette attached to $A$ or $B$, correspondingly. 
Next, take a path connecting vertices $A$ and $B$ by three links, as shown 
in Fig.\ref{connector}. Then the matrix $V_c$ entering the Bianchi identity 
for the cube $c$ is of the form
\begin{equation}
\label{ordprod}
V_{c} = \exp \left [ i\sigma^k\omega_k(c) \right ] = 
\left ( \prod_{p \in A} V_p \right ) \ C  
\left ( \prod_{p \in B} V_p \right ) \ C^{\dag} \ , \ 
C=U_{1} U_{2} U_{3}^{\dag} \ .
\end{equation}
$\prod_{p\in (A,B)}$ means an appropriately ordered product over three 
plaquettes of the cube attached to the vertex $A$ or $B$. 
The matrix $C$ defines a parallel transport of vertex $B$ into vertex $A$ 
and is called connector. 
The connector $C$ plays a crucial role in the nonabelian Bianchi identity. 
Its path can, in principle be chosen in an arbitrary way but the possible 
choices of $\prod_{p\in A}$, $\prod_{p\in B}$ and $C$ must fit together.
We choose the structure of connectors as shown in Fig.\ref{connector}. 
As is seen from Fig.\ref{connector} there are 
four types of elementary cubes with different types of connector pathes.
The next steps in the derivation are as follows:

I). Map one of the space-like plaquette delta's
to a delta-function for the cube $J(V_c)$ with $V_c$ 
given in (\ref{ordprod}).

II). Choose the maximal axial gauge in the form 
\begin{equation} 
\label{func5}
U_{3}(x,y,z)=U_{2}(x,y,0)=U_{1}(x,0,0)=I \ .
\end{equation}
\noindent
To simplify connectors one can fix the Dirichlet BC in the plane $z=0$.

III). Integrate out all link variables following \cite{plrepr}. 

In what follows we pass to the dual lattice where cubes become 
sites and so on. The partition function takes the form
\begin{equation}
\label{dualpart}
Z = \int \prod_{l}d V_{l}  \exp[\beta \sum_{l} 
{\rm Re \ Tr} V_{l}] \ \prod_{i=1}^4  \ \prod_{x(i)} \ 
J\left ( V_x^{(i)} \right ) \ ,
\end{equation}
\noindent
where $\prod_{x(i)}$ runs over all sites with $i$-th type of connector pathes 
and $J(V_x^{(i)})$ is an $SU(N)$ delta-function given by 
\begin{equation}
\label{func2}
J(V_x) =  \sum_r d_r \chi_r \left ( V_x \right ) \ ,
\end{equation}
\noindent
where the sum over $r$ is a sum over all representations of $SU(N)$,
$\chi_r$ is character of $r$-th representation and $d_r=\chi_r(I)$.

To write down exact expressions we choose coordinate 
system on the dual lattice as in Fig.\ref{connector}.
Then, e.g. the connector for the cube $B_1A$ is 
\begin{equation}
V_x^{(1)} =  V_{l_5}^{\dag}V_{l_1}V_{l_6}^{\dag} \ 
C_{\vec{x}(1)} \ V_{l_2}V_{l_3}V_{l_4}^{\dag} \ C_{\vec{x}(1)}^{\dag} \ ,
\label{V1}
\end{equation}
\noindent
\begin{equation}
C_{\vec{x}(1)} = \prod_{k=z_i-1}^1 V_{n_2}(x_i,y_i-1,k) 
\prod_{p=1}^{z_i-1}V_{n_1}(x_i-1,y_i,p) \ .
\label{CB1A}
\end{equation}
\noindent

\section{Low-temperature expansion}

In this section we construct the low-temperature 
expansion of gauge models in the plaquette formulation. In doing this 
we follow the strategy developed for the two-dimensional principal chiral models 
in \cite{2dsunlink}. The weak-coupling expansion in both models bears many 
common features. In the abelian case they turn out to be practically identical, 
while the only, but essential difference in the nonabelian case is the appearence 
of connectors in gauge models. 
The lattice PT in temporal gauge ($U_0=I$) was constructed 
for abelian and nonabelian models in the standard Wilson formulation 
by F.~M\"uller and W.~R\"uhl  \cite{pttemp}.  
In this gauge the PT for nonabelian models displays serious infrared 
divergences in separate terms of the expansion starting from the order  
${\cal O}(\beta^{-2})$ for Wilson loops and from 
${\cal O}(\beta^{-1})$ for the free energy. Although it is expected that 
all divergences are cancelled in gauge invariant quantities, we are not 
aware of any proof of the infrared finitness at higher orders of the expansion. 
Here we outline the low-temperature expansion for $SU(2)$ LGT
(all details and extension to other groups will be given in \cite{plaq}). 
We want to expand the partition and correlation functions
into asymptotic series whose coefficients are calculated over certain 
Gaussian measure. Let us consider the standard parameterization for 
the $SU(2)$ link matrix $V_l = \exp [i\sigma^k\omega_k(l)]$, 
where $\sigma^k, k=1,2,3$ are Pauli matrices and introduce
\begin{equation}
W_l = \left[ \sum_k\omega^2_k(l) \right]^{1/2} \ \ , \ 
W_x = \left[ \sum_k\omega^2_k(x) \right]^{1/2} ,
\label{Wp}
\end{equation}
\noindent
where $\omega_k(x)$ is a site angle defined in (\ref{ordprod})
and has the standard expansion $\omega_k(x) =  \omega^{(0)}_k(x) +
\omega^{(1)}_k(x) + \omega^{(2)}_k(x) + \cdots$ in powers of link angles.
The first coefficient can be written down as
\begin{equation}
\omega^{(0)}_k(x) = \omega_k(l_4) + \omega_k(l_5) + \omega_k(l_6) - 
\omega_k(l_1) - \omega_k(l_2) - \omega_k(l_3) \ .
\label{wkx0}
\end{equation}
\noindent
$\omega^{(i)}_k(x)$ for $i\geq 1$ can be computed by making repeated use 
of the Campbell-Baker-Hausdorff formula.
Then, the partition function (\ref{dualpart}) can be exactly rewritten 
to the following form \cite{2dsunlink}
\begin{eqnarray}
Z_{SU(2)} = \int \prod_l \left[ \frac{\sin^2W_l}{W^2_l}\prod_kd\omega_k(l) \right]
\exp \left[ 2\beta\sum_l\cos W_l \right] \prod_x \frac{W_x}{\sin W_x}
\nonumber     \\
\prod_x \sum_{m(x)=-\infty}^{\infty}\int\prod_kd\alpha_k(x)
\exp \left [ -i\sum_k\alpha_k(x)\omega_k(x) + 2\pi im(x)\alpha (x) \right ] \ ,
\label{PFwk}
\end{eqnarray}
\noindent
where $\alpha (x)=(\sum_k\alpha^2_k(x))^{1/2}$.
In derivation of this representation we have used the Poisson formula. 
In order to perform the weak coupling expansion we make the substitution
$\omega_k(l)\to (2\beta)^{-1/2}\omega_k(l)$, 
$\alpha_k(x)\to (2\beta)^{1/2}\alpha_k(x)$ and then expand the integrand 
of (\ref{PFwk}) in powers of fluctuations of the link fields.
Introduce external sources $h_k(l)$ coupled to the link field
$\omega_k(l)$ and $s_k(x)$ coupled to the auxiliary field $\alpha_k(x)$.
Then one can get the following expansion 
\begin{equation}
Z = \left [ 1+\sum_{k=1}^{\infty}\frac{1}{\beta^k}
B_k\left (\partial_h,\partial_s \right ) \right ] M(h,s) \ ,
\label{AppA1}
\end{equation}
where $B_k$ are known operators.
Up to exponentially small terms the generating functional $M(h,s)$ is 
given by 
\begin{equation}
\exp \left[ \frac{1}{4}s_k(x)G_{x,x^{\prime}}s_k(x^{\prime}) +
\frac{i}{2}s_k(x)D_l(x)h_k(l) +
\frac{1}{4}h_k(l)G_{ll^{\prime}}h_k(l^{\prime}) \right] \ ,
\label{GFfin}
\end{equation}
\noindent
where we introduced link Green functions $G_{ll^{\prime}}$ 
and $D_l(x^{\prime})$ defined as
\begin{equation}
G_{ll^{\prime}} = 2\delta_{l,l^{\prime}} - G_{x,x^{\prime}} -
G_{x+n,x^{\prime}+n^{\prime}} + G_{x,x^{\prime}+n^{\prime}} + 
G_{x+n,x^{\prime}} \ , 
\label{Gll1}
\end{equation}
\noindent
\begin{equation}
D_l(x^{\prime}) = G_{x,x^{\prime}} - G_{x+n,x^{\prime}} \ .
\label{Dxl}
\end{equation}
\noindent
These formulae allow us to calculate the weak coupling expansion of both 
the free energy and any fixed-distance observable. Extension of this expansion 
for $SU(N)$ LGT and for the Wilson loop and other observables is straightforward. 

Let us give some comments on the property of the expansion and on the 
infrared problem. One proves that the probability 
that $K$ links have large fluctuations like 
$\omega_k(l)\sim {\cal O}(\sqrt{\beta})$ is suppressed as 
${\cal O}(e^{-{\rm const}K\pi^2\beta})$, $C$ is a constant. 
It shows that large fluctuations are under good control since 
they are exponentially suppressed with $\beta$, and this obviously remains 
true in the TL. This property is very essential 
achievement of the expansion in the plaquette representation. No such 
controll exists for the large fluctuations of the link gauge matrices in 
the TL. It follows from the definition of the link 
Green function (\ref{Gll1}) that $\mid G_{ll^{\prime}} \mid \leq 4/3$. 
Hence, with respect to the Gaussian measure  
fluctuations of link variables are bounded like
\begin{equation}
\mid \langle \omega_l\omega_{l^{\prime}} \rangle_G \mid =
\mid \frac{1}{2}G_{ll^{\prime}}\mid \asymp  \frac{{\rm const}}{R^3} \ .  
\label{linkfluc}
\end{equation}
\noindent
This property justifies the low-temperature expansion in 
powers of fluctuations of link (plaquette) variables, while 
there is no such justification for the expansion in terms 
of the original link variables the fluctuations of which are not 
bounded when $L\to\infty$. 
As is seen from the generating functional (\ref{GFfin}) it depends only on 
infrared finite Green functions: link Green functions are infrared finite by 
construction, while the Green function for the free scalar 
field is finite in any dimension $D\geq 3$. 

In spite of all these nice properties, it is far from obvious that 
the expansion is uniformly valid in the volume. The problem appears in the 
following way. In \cite{plaq} we compute the first coefficient of the free 
energy expansion and prove that it coincides with the standard answer.   
Nevertheless, that computations show clearly that the source of the 
troubles are the connectors of the nonabelian Bianchi identities. 
The low-temperature expansion starts from the abelian Bianchi identity 
for every cube and goes towards gradual restoration 
of the full identity with every order of the expansion. Thus, the generating 
functional contains an abelianized form of the identity without connectors. 
High-order terms include an expansion of connectors which lead to appearence  
of infrared divergent terms. Consider, for example the term 
$\alpha_k(x)\omega_k^1(x)\sim {\cal O}(1/\sqrt{\beta})$ in the exponent of 
(\ref{PFwk}). $\omega_k^1(x)$ includes the following contributions from 
connectors 
$\epsilon_{kmn}(\omega_m(l_1)+\omega_m(l_2)+\omega_m(l_3))\sum_{l\in C} 
\omega_n(l)$, 
where sum over $l$ extends over all links belonging to the connector $C$.
For any finite $L$ one can always find such large $\beta$ that 
this sum is of the order ${\cal O}(1/\sqrt{\beta})$.  
In the weak coupling expansion such terms are treated as perturbations 
and should be expanded. These terms are not infrared finite. And even though 
the final result is infrared finite it is not obvious that such terms 
may be considered as perturbation in the TL. 
Indeed, in the infinite volume limit practically all connectors 
become infinitely long, therefore it is not clear if the last sum 
is properly bounded in this limit and may be expanded perturbatively.

\section{Dual representation, 't Hooft loop and monopoles}

In this section we give the dual form of $SU(2)$ LGT which can be obtained from 
the plaquette formulation.  
Let $x$ be a site dual to the cube of the original lattice and 
$l$ be a link dual to the plaquette of the original lattice. Then it follows from 
(\ref{dualpart}), (\ref{func2}) and from the expression for the Jacobian that in 
the case of the $SU(2)$ gauge group the partition function on the dual lattice 
can be written in the following form 
\begin{equation}
Z = \sum_{r_x=0,\frac{1}{2},1,... }^{\infty}\prod_x
\left [ (2r_x + 1) \sum_{m_i(x)=-r_x}^{r_x} \right ] \prod_l \Xi_l \ . 
\label{PFLsu2}
\end{equation}
\noindent
The summation over $r_x$ corresponds to the summation over all irreducible 
representations of the $SU(2)$ group. The sums over 
magnetic numbers $m_i(x)$ correspond to the calculation of $SU(2)$ 
traces. The index $i$ may run from $6$ to $6+3L$ depending on the position of 
the original cube, $L$ is the linear extent of the lattice.
The link integral $\Xi_l$ is given by 
\begin{equation}
\Xi_l = \int dV e^{\beta  {\rm Tr} V} \ 
V_{r_x}^{m_1n_1}V_{r_{x+n}}^{\dagger \ m_2n_2}
\prod_{i=1}^{M(x)}\left ( V_{r_i}^{k_ik_{i+1}}
V_{r_i}^{\dagger \ p_ip_{i+1}} \right ) \ ,
\label{linkint}
\end{equation}
\noindent  
where $V_{r}^{mn}$ is a matrix element of $r$-th representation.
The integer number $M(x)$ depends on the position of the plaquette 
on the lattice and shows how many times a dual link 
serves as connector in the Bianchi identities. 

Consider now the t' Hooft line which in $3D$ consists of a string $S_{xy}$ 
of dual links between sites $x$ and $y$ on the dual lattice. With the set 
$T$ of plaquettes dual to the string $S_{xy}$ we define the action 
\begin{equation}
S_z(U;T) = \beta\sum_{p\notin T}{\rm Re Tr}U_p + 
\beta\sum_{p\in T}{\rm Re Tr}zU_p 
\label{Taction}
\end{equation}
\noindent
and the corresponding partition function $Z(z;S_{xy})$ 
with action (\ref{Taction}). Here $z$ is center element of $SU(N)$. 
By a change of variables $V_l\to z^{*}V_l$ for $l\in S_{xy}$ 
we get the following representation for the partition function $Z(z;S_{xy})$ 
\begin{eqnarray}
Z(z;S_{xy}) = \int \prod_{l}d V_{l}  \exp[\beta \sum_{l} 
{\rm Re \ Tr} V_{l}] \ \prod_{i=1}^3  \ \prod_{x(i)} \ 
J\left ( V_x^{(i)} \right )  \nonumber  \\ 
\prod_{x(1)\ne x,y} \ J\left ( V_x^{(1)} \right ) \ 
J\left ( zV_x^{(1)} \right ) J\left ( zV_y^{(1)} \right ) \ ,
\label{SxyPF}
\end{eqnarray}
\noindent	
since $z$ cancels from all Bianchi constraints but those for the endpoints. 
Therefore, one can say that the Bianchi constraint is violated at these 
endpoints. It follows from these expressions that in the ensemble defined 
by the partition function (\ref{PFLsu2}) the disorder operator becomes 
particulary simple 
\begin{equation}
\langle D_z(S_{xy}) \rangle = \frac{Z(z;S_{xy})}{Z(1;S_{xy})} = 
\langle \exp [2i\pi (r_x-r_y)] \rangle \ . 
\label{disopdual}
\end{equation}
\noindent	 
Comparing this expression with the partition function (\ref{PFwk}) 
and remembering that $r_x$ is just the radial component of the 
auxiliary field $\alpha_k(x)$ we conclude that introducing 
sources like $r_x-r_y$ amounts to a simple shift in the corresponding 
sites for the integers $m_x$ in (\ref{PFwk}). 
Therefore, we may interpret the summation over $m_x$ as summation over 
monopole configurations which appear due to the periodicity of 
$SU(2)$ delta-function (in close analogy with $U(1)$ model). 
Indeed, expressed in terms of elements of algebra the Bianchi identity 
reads $( \sum_k\omega_k^2(x) )^{1/2} = 2\pi m_x$.
Configurations $m_x\ne 0$ we term monopoles of the maximal axial gauge. 
Investigation of these configurations and their contribution to 
the Wilson loops will be presented in a separate publication.

\end{document}